%% file: 2009_sidis.tex
\documentclass[a4paper]{cernart}
\usepackage{graphicx}
\graphicspath{{figures/}}
\usepackage{amssymb}
\usepackage{rotating}

\begin {document}
\dimen\footins=\textheight

%

\begin{titlepage}
\docnum{CERN--PH--EP/2009--008}
\date{9 April 2009}
\vspace{1cm}

\title{\LARGE
Flavour Separation of Helicity Distributions \\
from Deep Inelastic Muon--Deuteron Scattering}
\vspace*{0.5cm}

\collaboration{COMPASS Collaboration}


\vspace{2cm}
\input abstract.tex

\vspace*{60pt}


\vfill
\submitted{submitted to Physics Letters}

\noindent
{{\large  COMPASS Collaboration}\\[\baselineskip]}
\input{auth_cern.tex}%
%
\input{inst_cern.tex}\end{titlepage}

%


%
%
%






\input introduction.tex

\input spectrometer.tex
\input asym.tex

\input LO_fit.tex
\input sum_asym.tex

\input conclusion.tex

\input acknowledgement.tex

\input tables_withRC2.tex

\input bibliography.tex
\end{document}

%% file: abstract.tex
\begin{abstract}
We present a LO evaluation of helicity densities of valence, $\Delta u_v$$+$$\Delta d_v$,
non-strange sea, $\Delta\bar{u}$$+$$\Delta\bar{d}$, and strange quarks, 
$\Delta s$ (assumed to be equal to $ \Delta {\overline s}$).
They have been obtained from the  inclusive asymmetry $A_{1,d}$ and
the semi-inclusive asymmetries
$A^{\pi+}_{1,d}$, $A^{\pi-}_{1,d}$, $A^{K+}_{1,d}$, $A^{K-}_{1,d}$
measured in polarised deep inelastic muon-deuteron scattering.
The full deuteron statistics of COMPASS (years 2002--2004 and 2006) has been used.
The data cover the range $Q^2 > 1$\,(GeV/$c)^2$ and  $0.004<x<0.3$.
Both non-strange densities are found to be in a good agreement with previous 
measurements. The distribution of $\Delta s(x)$ is compatible with zero
in the whole measured range, in contrast to  the  shape of the strange quark
helicity distribution obtained in most LO and  NLO QCD fits.
The sensitivity of  the values of $\Delta s(x)$ 
upon the choice of fragmentation functions used in the derivation is discussed.


 
\end{abstract}

%% file: auth_cern.tex
%
%
%
M.~Alekseev\Iref{turin_p},
V.Yu.~Alexakhin\Iref{dubna},
Yu.~Alexandrov\Iref{moscowlpi},
G.D.~Alexeev\Iref{dubna},
A.~Amoroso\Iref{turin_u},
A.~Austregisilio\IIref{cern}{munichtu},
B.~Bade{\l}ek\Iref{warsaw},
F.~Balestra\Iref{turin_u},
J.~Ball\Iref{saclay},
J.~Barth\Iref{bonnpi}
% ...

%% file: inst_cern.tex
%
%
\Instfoot{bielefeld}{Universit\"at Bielefeld, Fakult\"at f\"ur Physik, 33501 Bielefeld, Germany\Aref{f}}
\Instfoot{bochum}{Universit\"at Bochum, Institut f\"ur Experimentalphysik, 44780 Bochum, Germany\Aref{f}}
\Instfoot{bonniskp}{Universit\"at Bonn, Helmholtz-Institut f\"ur  Strahlen- und Kernphysik, 53115 Bonn, Germany\Aref{f}}
\Instfoot{bonnpi}{Universit\"at Bonn, Physikalisches Institut, 53115 Bonn, Germany\Aref{f}}
\Instfoot{brno}{Institute of Scientific Instruments, AS CR, 61264 Brno, Czech Republic\Aref{g}}
\Instfoot{burdwan}{Burdwan University, Burdwan 713104, India\Aref{h}}
\Instfoot{calcutta}{Matrivani Institute of Experimental Research \& Education, Calcutta-700 030, India\Aref{i}}
\Instfoot{dubna}{Joint Institute for Nuclear Research, 141980 Dubna, Moscow region, Russia}
\Instfoot{erlangen}{Universit\"at Erlangen--N\"urnberg, Physikalisches Institut, 91054 Erlangen, Germany\Aref{f}}
\Instfoot{freiburg}{Universit\"at Freiburg, Physikalisches Institut, 79104 Freiburg, Germany\Aref{f}}
\Instfoot{cern}{CERN, 1211 Geneva 23, Switzerland}
\Instfoot{liberec}{Technical University in Liberec, 46117 Liberec, Czech Republic\Aref{g}}
\Instfoot{lisbon}{LIP, 1000-149 Lisbon, Portugal\Aref{j}}
\Instfoot{mainz}{Universit\"at Mainz, Institut f\"ur Kernphysik, 55099 Mainz, Germany\Aref{f}}
\Instfoot{miyazaki}{University of Miyazaki, Miyazaki 889-2192, Japan\Aref{k}}
\Instfoot{moscowlpi}{Lebedev Physical Institute, 119991 Moscow, Russia}
\Instfoot{munichlmu}{Ludwig-Maximilians-Universit\"at M\"unchen, Department f\"ur Physik, 80799 Munich, Germany\AAref{f}{l}}
\Instfoot{munichtu}{Technische Universit\"at M\"unchen, Physik Department, 85748 Garching, Germany\AAref{f}{l}}
\Instfoot{nagoya}{Nagoya University, 464 Nagoya, Japan\Aref{k}}
\Instfoot{praguecu}{Charles University, Faculty of Mathematics and Physics, 18000 Prague, Czech Republic\Aref{g}}
\Instfoot{praguectu}{Czech Technical University in Prague, 16636 Prague, Czech Republic\Aref{g}}
\Instfoot{protvino}{State Research Center of the Russian Federation, Institute for High Energy Physics, 142281 Protvino, Russia}
\Instfoot{saclay}{CEA DAPNIA/SPhN Saclay, 91191 Gif-sur-Yvette, France}
\Instfoot{telaviv}{Tel Aviv University, School of Physics and Astronomy, 69978 Tel Aviv, Israel\Aref{m}}
\Instfoot{triest_i}{Trieste Section of INFN, 34127 Trieste, Italy}
\Instfoot{triest}{University of Trieste, Department of Physics and Trieste Section of INFN, 34127 Trieste, Italy}
\Instfoot{triestictp}{Abdus Salam ICTP and Trieste Section of INFN, 34127 Trieste, Italy}
\Instfoot{turin_u}{University of Turin, Department of Physics and Torino Section of INFN, 10125 Turin, Italy}
\Instfoot{turin_i}{Torino Section of INFN, 10125 Turin, Italy}
\Instfoot{turin_p}{University of Eastern Piedmont, 1500 Alessandria,  and Torino Section of INFN, 10125 Turin, Italy}
\Instfoot{warsaw}{So{\l}tan Institute for Nuclear Studies and University of Warsaw, 00-681 Warsaw, Poland\Aref{n} }
\Instfoot{warsawtu}{Warsaw University of Technology, Institute of Radioelectronics, 00-665 Warsaw, Poland\Aref{o} }
\Instfoot{yamagata}{Yamagata University, Yamagata, 992-8510 Japan\Aref{k} }
%
%
\Anotfoot{+}{Deceased}
\Anotfoot{a}{Also at IST, Universidade T\'ecnica de Lisboa, Lisbon, Portugal}
\Anotfoot{b}{On leave of absence from JINR Dubna}
\Anotfoot{c}{Also at Chubu University, Kasugai, Aichi, 487-8501 Japan$^{\rm j)}$}
\Anotfoot{d}{Also at KEK, 1-1 Oho, Tsukuba, Ibaraki, 305-0801 Japan}
\Anotfoot{e}{Also at GSI mbH, Planckstr.\ 1, D-64291 Darmstadt, Germany}
\Anotfoot{f}{Supported by the German Bundesministerium f\"ur Bildung und Forschung}
\Anotfoot{g}{Suppported by Czech Republic MEYS grants ME492 and LA242}
\Anotfoot{h}{Supported by DST-FIST II grants, Govt. of India}
\Anotfoot{i}{Supported by  the Shailabala Biswas Education Trust}
\Anotfoot{j}{Supported by the Portuguese FCT - Funda\c{c}\~ao para a Ci\^encia e Tecnologia grants POCTI/FNU/49501/2002 and POCTI/FNU/50192/2003}
\Anotfoot{k}{Supported by the MEXT and the JSPS under the Grants No.18002006, No.20540299 and No.18540281; Daiko Foundation and Yamada Foundation}
\Anotfoot{l}{Supported by the DFG cluster of excellence `Origin and Structure of the Universe' (www.universe-cluster.de)}
\Anotfoot{m}{Supported by the Israel Science Foundation, founded by the Israel Academy of Sciences and Humanities}
\Anotfoot{n}{Supported by Ministry of Science and Higher Education grant 41/N-CERN/2007/0}
\Anotfoot{o}{Supported by KBN grant nr 134/E-365/SPUB-M/CERN/P-03/DZ299/2000}

%% file: introduction.tex
\section {Introduction}

Among the various sea quarks contributing to the nucleon spin,
the strange quark is the only one accessible in
inclusive lepton-nucleon scattering experiments. The first moment
of the strange quark helicity distribution,
$\Delta s$$+$$\Delta {\overline s}$, has been found to be negative 
already twenty years ago in the EMC experiment \cite{emc} 
under the assumption of SU(3)$_F$ symmetry in hyperon $\beta$ decays. 
This result has been confirmed with improved precision by recent 
measurements performed by HERMES \cite{hrm2007}
and by COMPASS which has obtained
\begin{eqnarray}
\Delta s + \Delta {\overline s} = -0.09 \pm 0.01 {\rm(stat.)} \pm 0.02 {\rm(syst.)}
\label{mom_cmp}
\end{eqnarray}
at $Q^2 = 10 ({\rm GeV}/c)^2$
at leading order (LO) in  QCD \cite{dval}.
Inclusive experiments, however,  provide an evaluation of the first moment
$\Delta s$$+$$\Delta {\overline s}$ only.
The shape as a function of the Bjorken scaling variable $x$ 
is determined in  global fits of the nucleon spin structure 
function $g_1(x,Q^2)$ where a parameterisation of 
the strange quark helicity as a function of $x$  assumed to be 
valid at some reference value of the photon virtuality, $Q_0^2$, is evolved 
to the  $Q^2$ of each data point and fitted to the measured values.
The resulting distribution  $\Delta s(x)$, further assumed to be equal to
$\Delta {\overline s}(x)$, is generally concentrated at
the highest values of $x$ compatible with the positivity limit
$|\Delta s(x,Q^2)| \le s(x,Q^2)$ \cite{g1d2006,LSS}.

Direct information on the distribution $ \Delta s(x)$ can be
obtained from semi-inclusive channels, in which  interactions on 
strange quarks are enhanced, such as charged kaon production. These
measurements, which require final state particle identification, 
became only feasible in recent experiments and the only results published 
so far are from the HERMES experiment \cite{hrm2004,hrm2005,hrm2008}.

In a  full flavour decomposition analysis, the HERMES collaboration
has obtained $\Delta s = 0.028 \pm  0.033({\rm stat.}) \pm 0.009({\rm syst.})$
in the range  $0.023 <x< 0.6$ \cite{hrm2005}.
In their most recent  analysis of the charged kaon and inclusive asymmetries
in  deuteron data,
they have obtained
$\Delta s+\Delta {\overline s} = 0.037\pm0.019({\rm stat.})\pm0.027({\rm syst.})$
 \cite{hrm2008}. The negative values of $\Delta s(x)$ expected from the
full first moment have thus never been observed in the $x$ range covered by HERMES.
The implication of a positive $\Delta s$  in a limited
experimental  range 
was discussed 
in  relation with the assumed $SU(3)$ flavour symmetry and it was shown 
that a non-negative first moment of $\Delta s$ was highly unlikely \cite{leader-sta}. 
This situation is clearly reflected in the result of a global fit
 including all inclusive and semi-inclusive results
in the DIS region:  the fitted distribution
of  $\Delta s(x)$ which is positive at $x > 0.03$
  receives  a negative contribution in the unmeasured
low $x$ range, to bring its first moment close to the values
of Eq.\,(\ref{mom_cmp}) \cite{florian2008}.

 In this paper we present a new  precise measurement of the inclusive and
semi-inclusive double spin asymmetries measured
on an isoscalar target by  the COMPASS experiment at CERN
and a LO evaluation of the polarised parton
distributions $\Delta u_v$$+$$\Delta d_v$, $\Delta {\overline u}$$+$$\Delta {\overline d}$
and $\Delta s$ (=$\Delta {\overline s})$.

At  LO in QCD under the assumption of independent quark fragmentation,
the double spin asymmetries for hadrons $h$ produced in the
current fragmentation region can be decomposed into a sum of products of quark 
helicity distributions $\Delta q(x,Q^2)$ times quark fragmentation
 functions~$D_q^h(z,Q^2)$ where $z$ is the fraction of the virtual photon
energy taken by the hadron $h$: 
\begin{eqnarray}
A^h_1(x,Q^2,z) = \frac{\sum_q e_q^2 \Delta q(x,Q^2) D_q^h(z,Q^2)}{\sum_q e_q^2 q(x,Q^2) D_q^h(z,Q^2)} \,.
\label{indep_frag}
\end{eqnarray}

A previous  determination of  the quark helicity distributions  performed
by  SMC \cite{SMC_98} covers a kinematic range similar to the COMPASS data 
but does not provide a determination of $\Delta s$ due to the lack of 
hadron identification.

The  deuteron data presented in this paper were collected
in the years 2002-2004 and 2006. The produced hadrons cover a large
phase space. 
In the present analysis only those identified as pions or kaons are used.

%% file: spectrometer.tex
\section{Experimental setup}

A general description of the COMPASS spectrometer in the initial configuration
is given in \cite{spectro}.
Only modifications introduced after the year 2005 will be mentioned here.
They  mainly concern the polarised
target, the large area trackers around the
 first spectrometer  magnet (SM1) , and the RICH detector.

\begin{figure}
\includegraphics[width=0.7\hsize]{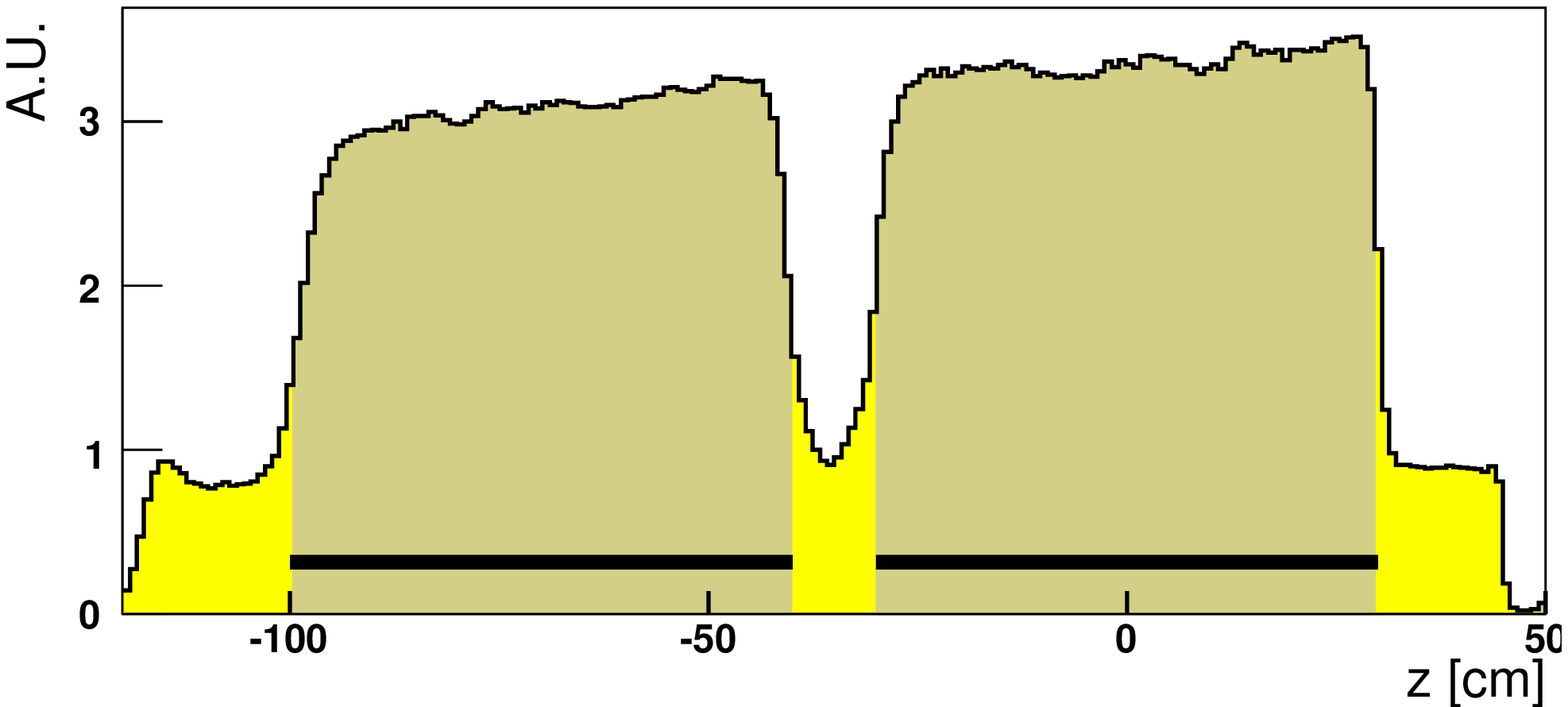}\\
\includegraphics[width=0.7\hsize,clip]{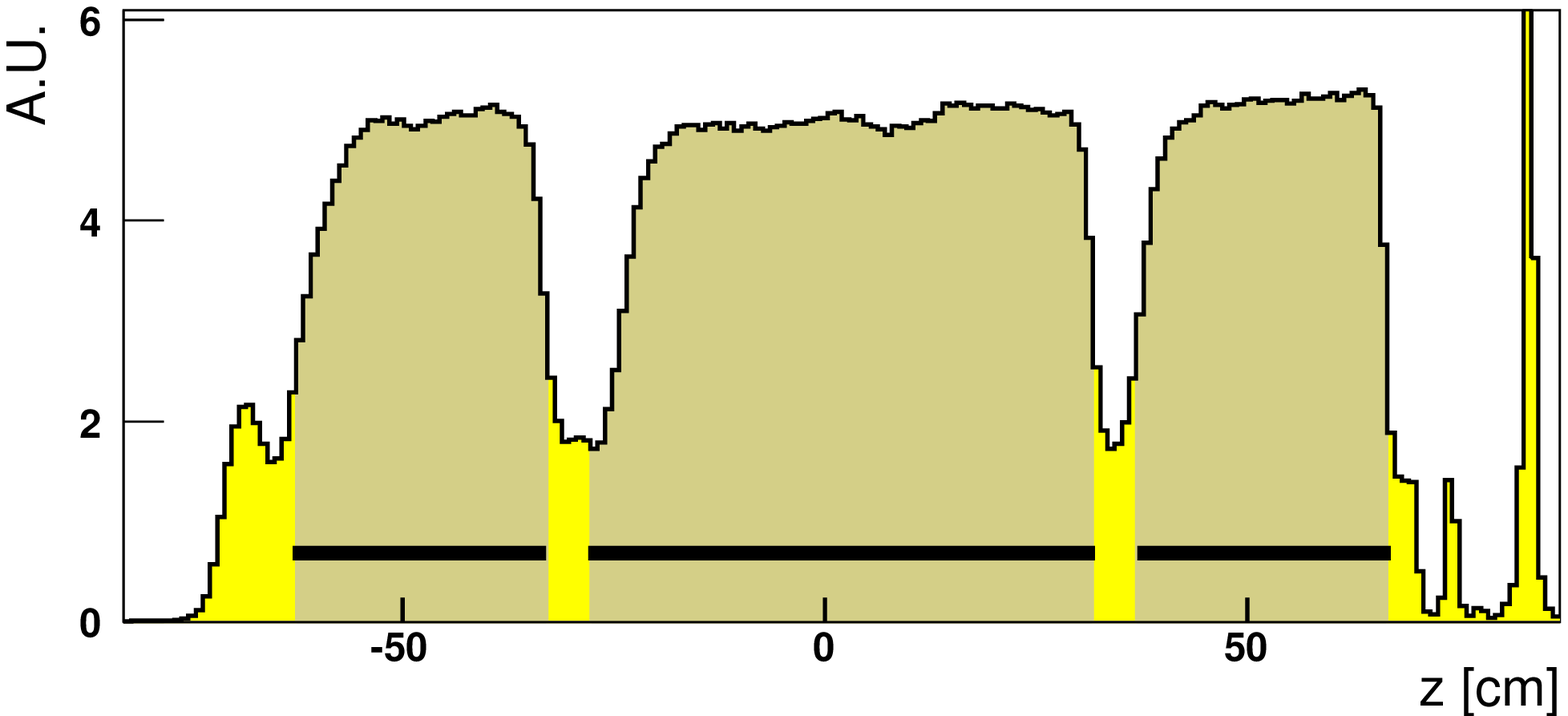}
\caption{
   Distribution of the interaction vertices along the beam axis
  for the 2 cell
  (years 2002--2004) and 3 cell (year 2006) target configurations. 
  The solid lines show the length of the cells.
}
\label{fig:z_vert}
\end{figure}

Before 2005 the target solenoid magnet was  the one previously used by 
the SMC experiment \cite{smc_ta} with an aperture of $\pm 70$\,mrad as seen 
from the upstream end of the target. 
The new solenoid magnet \cite{new_target} installed in 2005 has an acceptance 
of $\pm 180$\,mrad.
Before 2005 the polarised target,  located inside the solenoid, consisted of two cells, 
each 60\,cm long and 3\,cm in diameter, separated by 10\,cm.  
A three cell target has been installed in the new magnet.
In this configuration the central cell is 60\,cm long and the two outer ones 
30\,cm long, separated by 5\,cm. The total amount of material thus remains unchanged.
The distribution of the interaction vertices along the beam axis
for the events used in the present analysis 
is shown in Fig.\,\ref{fig:z_vert} for the old and new target configurations.

The deuterated lithium target material ($^6$LiD) is longitudinally
polarised with the method of dynamic nuclear polarisation (DNP)\,\cite{target}.
In the old as well as in the new configuration neighbouring  cells of 
the targets are polarised in opposite directions so that data from both spin 
directions are recorded at the same time.    
The absolute value of the averaged polarisation varies between $0.50$ and $0.56$.
Before 2005 the spin directions in the target cells were reversed every 8 hours
by rotating the magnetic field direction.               
In this way, fluxes and acceptances cancel out in the calculation of spin asymmetries, 
provided that the ratio of acceptances remains unchanged after spin reversal.
In the new configuration, the data samples obtained with both spin orientations
have in average the same acceptance, which limits false asymmetries.
In view of this, the magnetic field direction was rotated only once per day during
the 2006 data taking.
In order to minimise possible acceptance effects related to the orientation of 
the solenoid field, the sign of the polarisation in each target cell was also 
reversed a few  times per year by changing the DNP microwave frequencies.

Several modifications have been introduced in the tracking detectors around 
SM1 in order to match the enlarged acceptance of the new solenoid: 
an additional medium size drift chamber station (DC) has been installed upstream 
of SM1 and two smaller DCs, downstream of SM1, have been replaced by a new larger one 
and by a  straw tube station.

A major upgrade has also been applied to the RICH detector  
to improve its performance in terms of efficiency and purity\,\cite{rich_upgd}:
in the most critical central region, photon detection previously provided by large-size
MWPCs with CsI photocathodes 
has been replaced by   
a   system based on multi-anode PMTs.
It improves considerably the signal-to-noise ratio in the region
where the beam halo is largest.
In addition the readout system for the peripheral region has been 
replaced by a  faster one.

%
%

%% file: asym.tex
\section{Asymmetries}


All events used in the present analysis are required to have a reconstructed primary
interaction vertex defined initially by the incoming and the scattered muon trajectories
(the reconstruction procedure is described in Ref.\cite{tracking}).
The energy of the beam muon is constrained to be in the interval $140<E_\mu<180$ GeV.
To equalise fluxes through the different target cells, it is required for
the trajectory of the incoming muon to cross entirely all cells. 
This condition is essential
 because it allows to cancel out the muon flux in the calculation of asymmetries.
The kinematic region is defined by cuts on the photon virtuality, $Q^2$,
and the fractional energy, $y$, transfered from the beam muon to the virtual photon.
 DIS events are selected by requiring $Q^2>1$\,(GeV/$c)^2$. The 
 requirement $y>0.1$
removes events affected by bad resolution  and low photon polarisation.
The region  most affected by radiative
corrections is eliminated by the cut
 $y<0.9$.

The $x$ range covered in the present analysis 
extends from 0.004 to 0.3. The low limit is defined by the 
kinematical cut  $Q^2>1$\,(GeV/$c)^2$. 
The upper limit is set to $x = 0.3$ because interactions on
 sea quarks 
are  negligible at higher $x$.
The total energy of the $\gamma^{*}$-nucleon system  in the selected events covers the
range $5 \lesssim W \lesssim 17$ GeV.
The statistics of 135.1 million events includes 
the sample used in  Ref.\cite{g1d2006}  
and reduces the statistical errors
by about 30 \%.

Hadron tracks are required to originate from the main vertex.
A cut on the  fractional energy $z > 0.2$ is applied 
to the hadron candidates in order to select those
produced in the current fragmentation region.
In addition an upper limit $z < 0.85$ is imposed in order to suppress hadrons from
diffractive processes and to avoid contamination from wrongly identified
muons. Several other cuts are applied to guarantee the quality of the selected
track sample: the first reconstructed track point  must be upstream of SM1;
 tracks reconstructed 
 only upstream of SM1 are rejected and those
crossing more than $30$ radiation lengths
of material are not accepted as hadrons.

Hadron identification is performed using the RICH detector.
For the present analysis, the momenta of hadrons are restricted 
to the range common to pion and kaon identification $10<p<50$ GeV$/c$.
The expected distributions of photo-electrons are calculated for different
particle masses as well as for the background assumption. These distributions
are compared to the observed one and the mass is assigned according to the
ratios of their likelihoods.
The  statistics available for the  $\pi^+$($\pi^-$) and $K^+$($K^-$)
samples after all cuts is  22.8(20.5), and 4.8(3.3) millions, respectively.


Since the samples of identified pions and kaons 
do not fully correspond to the true ones,
an unfolding procedure must be applied to correct rates and asymmetries.
In a first step, the elements of the identification 
efficiency matrix, $P^{t\rightarrow i}$, are calculated. They represent the 
probability for a particle of  true type $t$ to be identified as  type $i$.
The values of $P^{t\rightarrow i}$ are obtained from samples  of    
reconstructed pions and kaons resulting from  $K_S^0$ and $\phi$ decays, respectively.
Since the RICH performance depends critically on the phase space of particles,
the elements $P^{t\rightarrow i}$ are calculated in bins of momentum and polar 
angle (angle w.r.t. beam axis) of the selected particle.

%

\begin{figure}[tb]
\centering
\includegraphics[width=0.7\hsize,clip]{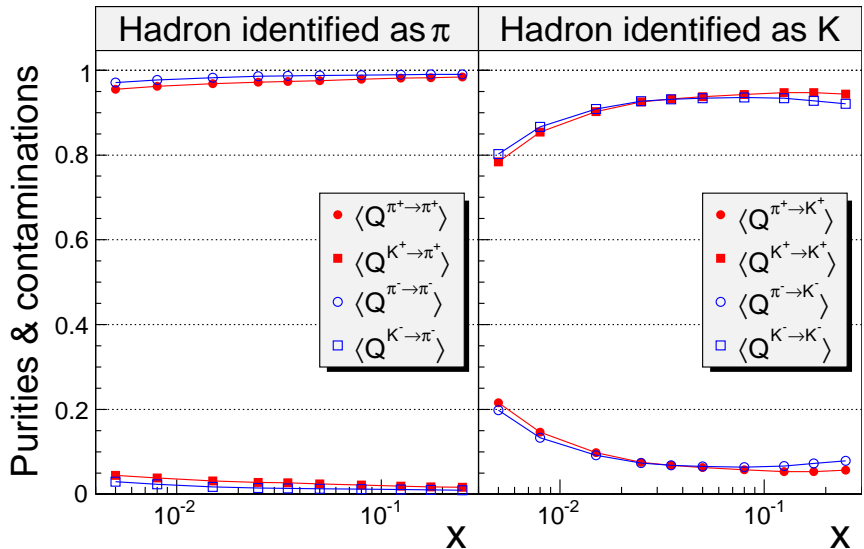}
\caption {Purities  $Q^{{\pi}\rightarrow {\pi}}$, $Q^{K\rightarrow K}$
 and contaminations  $Q^{{\pi}\rightarrow K}$, $Q^{K\rightarrow {\pi}}$ calculated for 
  the 2004 data.}
\label{fig:purity}
\end{figure}

 In  a second step the contributions , $Q^{t\rightarrow i}$, from different hadron
species $t$ to the identified sample $i$ are determined. They depend not
only on the identification efficiencies $P^{t\rightarrow i}$ but also on the
observed
hadron rates and therefore must be calculated separately for every bin of $x$.  
As an example the contributions $Q^{\pi \rightarrow \pi}$, $Q^{K \rightarrow K}$
(called "purities") and $Q^{\pi \rightarrow K}$, $Q^{K \rightarrow \pi}$
(called "contaminations") for positive and negative hadrons in the 2004 data  
 are shown in Fig.\,\ref{fig:purity} as a function of $x$.
In general, $Q^{{\pi}\rightarrow {\pi}}$ is close to 1.0 and $Q^{K\rightarrow K}$
varies from about 0.8 at low $x$ to about 0.93 at medium $x$.
In view of this, unfolding can only have a significant effect at low $x$
but since the
 pion and kaon asymmetries are similar in this region, its effect
remains small as compared to the 
statistical errors.



\begin{figure*}[tb]
\centering
\includegraphics[width=0.8\hsize,clip]{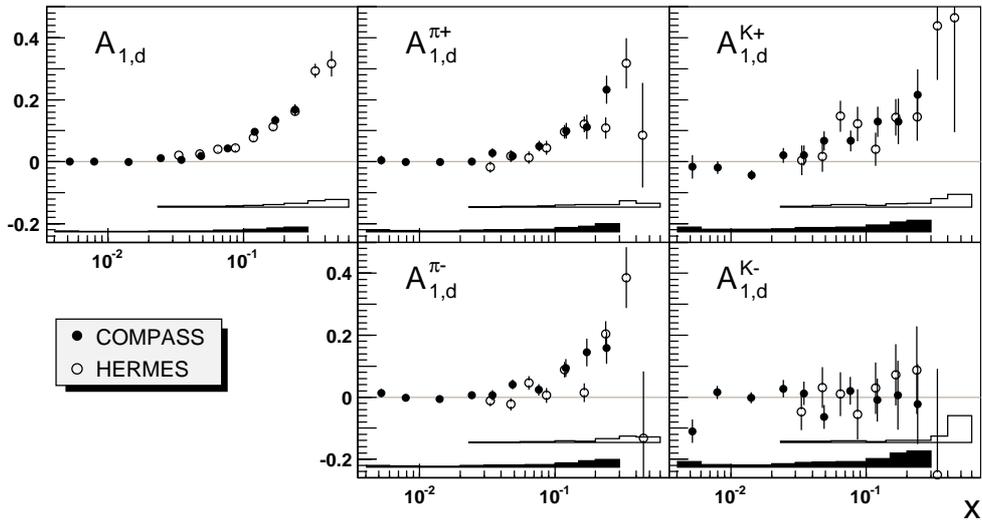}
\caption{Comparison of final asymmetries of COMPASS 
as a function of $x$ with results of HERMES \cite{hrm2005}.
  Bands at bottom of graphs represent systematic uncertainties.
  Solid markers and bands correspond to COMPASS data.
  Open markers and bands are taken from the HERMES publication.
}
\label{fig:akpmhpm}
\end{figure*}

The inclusive  asymmetries and the unfolded  hadron asymmetries 
have been corrected for radiative effects according
to the procedure of Ref.\cite{Shumeiko} and
are shown in Fig.\,\ref{fig:akpmhpm} as a function of $x$.
The  values of the inclusive asymmetry
 are in good agreement with those of Ref.\cite{g1d2006}.
The results of HERMES, the only other experiment which measured asymmetries of 
identified hadrons \cite{hrm2005}, are shown for comparison.
The two sets of measurements  are well compatible and in the region of kinematic
overlap the statistical precision  of the two experiments is generally comparable.
However COMPASS extends the measured region towards $x=0$ by an order of magnitude.
It is also observed that all asymmetries, except $A^{K-}_{1,d}$, are quite similar
to each other. This feature is expected due 
to the  isoscalar symmetry of the $^6$LiD target.
The $K^-$ asymmetry is consistent with zero over the full range of $x$.
The values of the inclusive and semi-inclusive asymmetries are listed with
their statistical and systematic errors in Tables \ref{tab:DIS_asym} and \ref{tab:SIDIS_asym}.
Correlations between different asymmetries in bins of $x$ are listed in Table \ref{tab:corr}.

There are several sources of systematics uncertainties in the determination
of the asymmetries. The error of the target polarisation measurement and 
the error on the parameterisation
of the beam polarisation amount to 5\%  of their respective value. 
 The uncertainty related to the
dilution factor, 
which includes the dilution due to radiative events on the deuteron,
is 2\% over the full range of $x$. The ratio
$R=\sigma^L/\sigma^T$ used to calculate the depolarisation factor \cite{abe} gives
an error of 2-3\%. When  added in quadrature these multiplicative uncertainties 
amount to a systematic error of 8\%  of the asymmetry.
The systematic error also accounts for false asymmetries
 which could be generated by instabilities in some
components of the spectrometer. 
Asymmetries due to apparatus effects have been searched for in combinations of
data samples where the physical asymmetry cancels out. 
The asymmetries observed in these combinations
 were found compatible with zero.  Systematic effects have
also been studied by comparing results obtained with different microwave
settings. No significant difference was found.
The possible error due to false asymmetries was evaluated as a fraction of 
the statistical error: $\sigma_{syst}\!<0.4\, \sigma_{stat}$\cite{g1d2006}. The total systematic
uncertainty is shown by the bands at the bottom of 
each plot in Fig.\,\ref{fig:akpmhpm}.

%% file: LO_fit.tex
\section{Polarised PDFs from a fit to the asymmetries}

As in our previous LO analysis \cite{dval}, we assumed that hadrons in the current
fragmentation region are produced by independent quark fragmentation,
so that their spin asymmetries can be written in terms of parton
distribution functions ($q(x,Q^2),\Delta q(x,Q^2)$) and fragmentation
functions (FFs) $D_q^h(z,Q^2)$ according to Eq.\,(\ref{indep_frag}). In the present analysis we
use the unpolarised parton distribution functions 
(PDFs) from MRST \cite{MRST}\footnote{We use the LO set with three quark 
flavours.}
 and 
the recent DSS parameterisation of FFs at LO which
 was obtained from a combined analysis of inclusive pion and kaon production data
from $e^+e^-$ annihilation, semi-inclusive DIS data ("SIDIS") 
from HERMES and proton-proton collider data \cite{florianFF}.
In order to test the dependence of the polarised PDFs on the FFs, we also show
the values obtained with the EMC FFs  \cite{EMC_FF}. In
contrast to other parameterisations which are derived from global fits, the
latter ones have been  extracted from the EMC data only, so that only  
non-strange quark fragmentation could be measured. 
Therefore, in addition to the 
assumptions  generally made to reduce the
number of FFs, the EMC analysis also assumed that $D_{\overline s}^{K+} =  D_u^{\pi+}$.

The recent HKNS parameterisation of FFs \cite{HKNS} strongly disagrees
with the ratio of negative to positive hadrons observed in our data,
as was already observed in \cite{dval} for the KRE parameterisation \cite{KRE}.
For this reason, these parameterisations based only on $e^+e^-$ collider data
are not usable in the kinematic range of the present analysis. 

Since the analysis is based on deuteron data only, 
only the sums  of $u$ and $d$ densities can be extracted:
$\Delta u_v$$+$$\Delta d_v$ and $\Delta\bar{u}$$+$$\Delta\bar{d}$.
In principle $\Delta{s}$ and $\Delta\bar{s}$ could both be extracted from 
the charged kaon asymmetries $A^{K+}_{1,d}$ and $A^{K-}_{1,d}$ but in view of 
the precision of the data, they are assumed to be equal.
All asymmetries are also assumed to be independent of $Q^2$.
In this way  the resulting PDFs are obtained at  
a common $Q^2$ fixed to 3\,(GeV/$c)^2$.

The five measured asymmetries form a linear system 
of equations with three unknowns
($\Delta u_v$$+$$\Delta d_v$, $\Delta\bar{u}$$+$$\Delta\bar{d}$, $\Delta s$),
which is solved by a  least-square fit independently in each $x$-bin.
Only statistical errors are used in the fit and correlations between
asymmetries are taken into account.
Two  corrections $(c_1,c_2)$ 
 are  applied 
in the evaluation of quark helicity distributions from the  asymmetries. 
The first one, $c_1 = 1 - 1.5 \omega_D$, 
accounts for the deuteron D-state contribution ($\omega_D = 0.05\pm0.01$ \cite{mch}).
The second one 
accounts for the fact that, although $R(x,Q^2) = 0$ at LO, 
the unpolarised PDFs originate from $F_2$ distributions 
in which $R = \sigma_L/\sigma_T$ was different from zero \cite{abe}.
In the present analysis we assume  $R$ to be the same for inclusive and semi-inclusive
reactions,
so that the same correction, $c_2 = 1 + R(x,Q^2)$,  can be used for
inclusive and hadron asymmetries. The resulting quark helicities thus are
corrected by a factor $\xi = c_1 \cdot c_2$.
 

\begin{figure}[tb]
\centering
\includegraphics[width=0.8\hsize,clip]{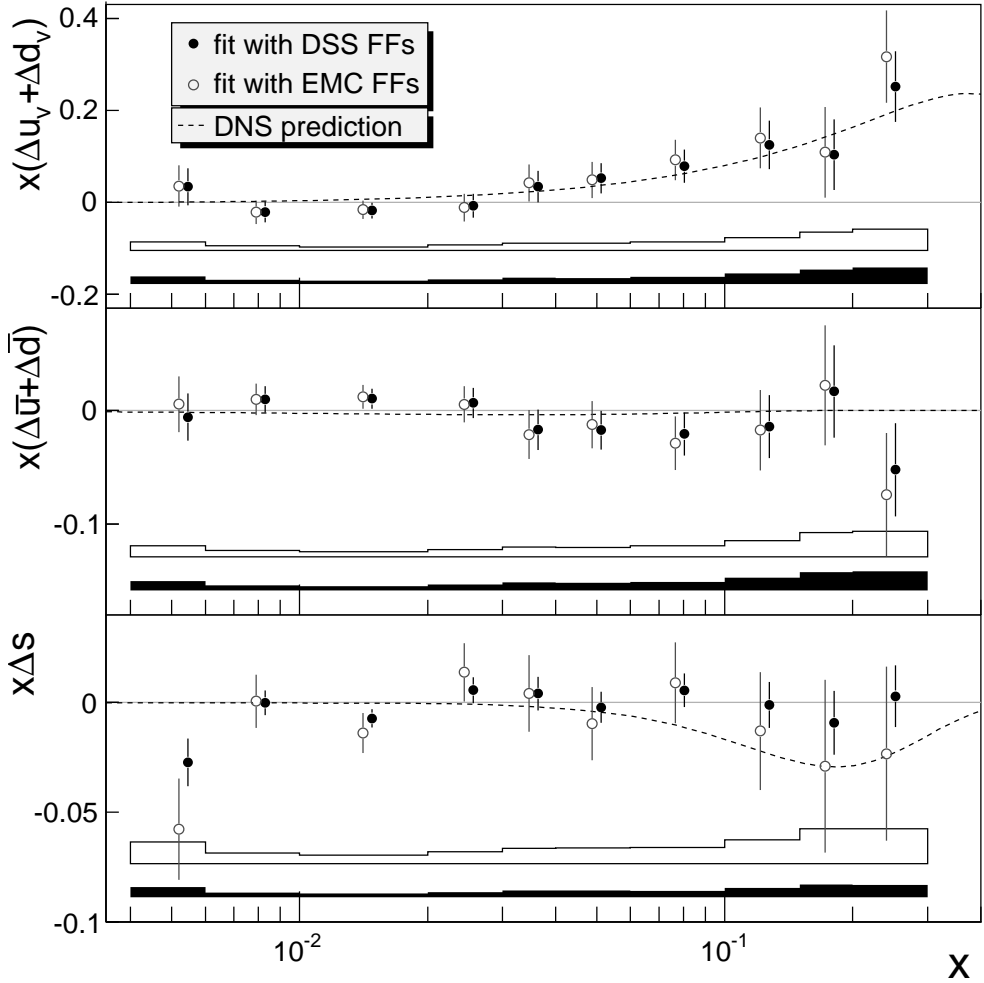}
\caption{
  The quark helicity distributions evaluated at common value $Q^2=3$\,(GeV/$c)^2$
  as a function of $x$ for two sets of fragmentation functions (DSS and EMC).
  Bands at bottom of graphs represent systematic uncertainties.
  Solid markers and bands correspond to PDFs obtained with DSS parameterisation of FFs.
  Open markers and bands are obtained with EMC parameterisation of FFs.
  The curves represent the LO DNS parameterisation of polarised PDFs \cite{DNS}.
}
\label{fig:PDF_Syst_EMC_DNS}
\end{figure}
\begin{figure}[tb]
\centering
\includegraphics[width=0.7\hsize]{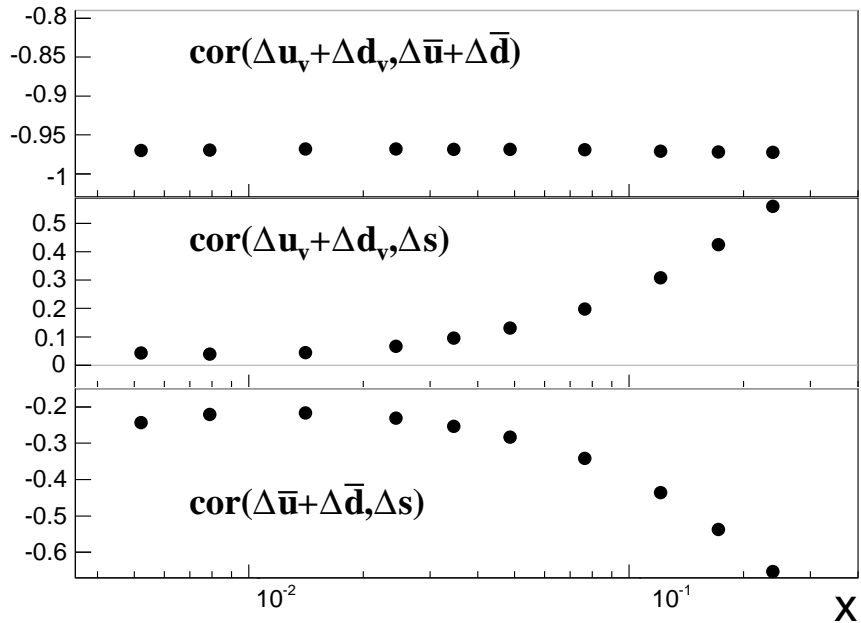}
\caption{
  Correlation coefficients of PDFs obtained in the fit with DSS parameterisation
  as a function of $x$.
}
\label{fig:PDF_cor}
\end{figure}

The results of the fit obtained with the two sets of fragmentation functions
are shown in Fig.\,\ref{fig:PDF_Syst_EMC_DNS}. 
Significant differences  are observed only for $\Delta s$.
Indeed the main difference of DSS with respect to EMC is the enhanced
$s$($\bar{s}$) quark contribution to the production of $K^-$($K^+$):
the ratio $\int_{0.2}^1 D_{\bar{s}}^{K+}(z)dz/\int_{0.2}^1D_u^{K+}(z)dz$, which 
is equal to 3.4 for the quoted EMC values, increases to 6.6 in DSS. The statistical
precision of $\Delta s$ for the two parameterisations changes in the same proportion.
The curves obtained with the LO DNS  parameterisation of polarised PDFs \cite{DNS} 
are also shown.
As in our previous publication on the asymmetry of unidentified hadrons \cite{dval},
a nice agreement is observed in the valence sector.
The asymmetries of the non-strange sea are also compatible with the DNS curve,
although we observe a tendency for the data points to be above and below the curve
at low and high $x$, respectively.
The shape of the $x\Delta s$ curve of DNS is quite typical for QCD fits of $g_1(x,Q^2)$ data,
showing a  minimum in the medium $x$ region $(x \approx 0.2)$.
With the DSS fragmentation functions,
the SIDIS measurements of  COMPASS do not seem to support this behaviour,
while with the EMC ones, the errors become too large to draw any conclusion.

The elements of the correlation matrix for the obtained densities are shown in Fig.\,\ref{fig:PDF_cor}.
The correlation between the non-strange densities $\Delta u_v$$+$$\Delta d_v$
and $\Delta\bar{u}$$+$$\Delta\bar{d}$ is large and negative.
This feature can be explained by the fact that their sum is highly
constrained by the very precise value of $A_{1,d}$: 
since the term with $\Delta s$ in Eq.\,(\ref{indep_frag}) is smaller than 
the other ones, $A_{1,d}$ fixes well the sum of non-strange densities and 
forces them to anti-correlate.

The estimates of the truncated first moments
 $\Delta u_v$$+$$\Delta d_v$, $\Delta\bar{u}$$+$$\Delta\bar{d}$
and $\Delta s$  are given in Table\,\ref{tab:FirstMom}.
The systematic errors have been estimated by refitting the
asymmetries shifted simultaneously within the limits of their
systematic uncertainty.
The value quoted for valence  quarks is in good agreement 
with the one derived in our previous publication 
from the  difference asymmetries for non-identified hadrons
 obtained from a partially overlapping data sample 
($0.26 \pm 0.07 \pm 0.04$ at $Q^2$$=$$10$\,(GeV/$c)^2$)~\cite{dval}.

\begin{table}[t]
\caption{  First moments $\Delta u_v$$+$$\Delta d_v$, $\Delta\bar{u}$$+$$\Delta\bar{d}$ 
  and $\Delta s$ at $Q^2=3$\,(GeV/$c)^2$ 
from the COMPASS data and also from the DNS fit at LO \cite{DNS} truncated to
  the range of the measurements ($0.004 < x < 0.3$).}
\label{tab:FirstMom}

\vspace*{3mm}
{
\begin{tabular}{|c|c||c|c|}
\hline
\multicolumn{2}{|c||}{FF} & DSS & EMC \\
\hline
\hline
\multicolumn{1}{|c|}{}& $\!\!$measur.$\!\!$ & $0.26\pm0.06\pm0.02\!\!$ & $0.30\pm0.08\pm0.02\!\!$ \\
\cline{2-4}
\multicolumn{1}{|c|}{\raisebox{1.5ex}[-1.5ex]{$\!\!\Delta u_v$$+$$\Delta d_v\!\!\!$}} & DNS & \multicolumn{2}{|c|}{0.225} \\
\hline
\hline
\multicolumn{1}{|c|}{}& $\!\!$measur.$\!\!$ & $\!\!-0.04\pm0.03\pm0.01\!\!$ & $\!\!-0.05\pm0.04\pm0.01\!\!$ \\
\cline{2-4}
\multicolumn{1}{|c|}{\raisebox{1.5ex}[-1.5ex]{$\Delta\bar{u}$$+$$\Delta\bar{d}$}}&DNS&\multicolumn{2}{|c|}{$-0.009$} \\
\hline
\hline
\multicolumn{1}{|c|}{}& $\!\!$measur.$\!\!$ & $\!\!-0.01\pm0.01\pm0.02\!\!$ & $\!\!-0.05\pm0.03\pm0.03\!\!$ \\
\cline{2-4}
\multicolumn{1}{|c|}{\raisebox{1.5ex}[-1.5ex]{$\Delta s$(=$\Delta\bar{s}$)}} & DNS &\multicolumn{2}{|c|}{$-0.035$}  \\
\hline
\end{tabular}
}
\end{table}

%% file: sum_asym.tex
\section {Direct evaluation of $\Delta s$ from the charged kaon asymmetry}

\begin{figure}[tb]
\begin{center}
\includegraphics[width=0.9\hsize,clip]{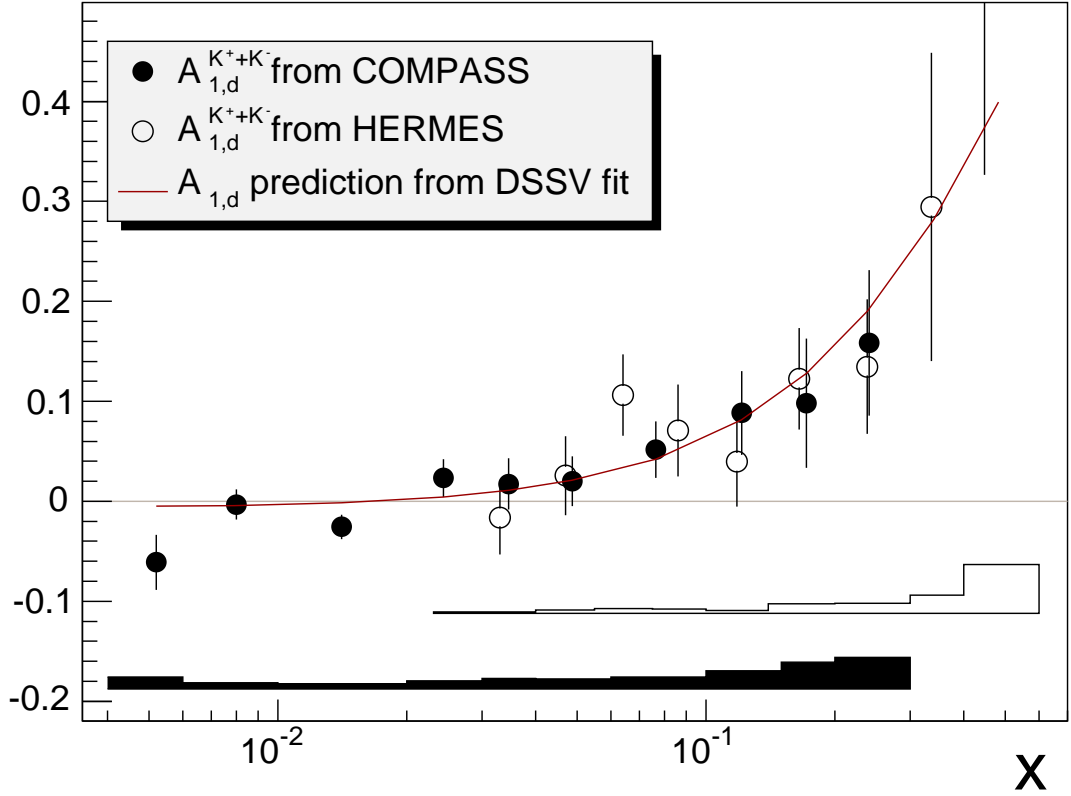}
\end{center}
\caption{Charged kaon asymmetries obtained with cross-section weights from
  MRST PDFs and DSS FFs. For comparison results of HERMES \cite{hrm2005}
  are also presented. The curve shows the $A_{1,d}$ prediction of the
  DSSV fit \cite{florian2008}.
}
\label{fig:akpmchi}
\end{figure}

The dependence of $\Delta s(x)$ on the FFs can be further explored in relation
with the charged kaon asymmetry $A^{K^++K^-}_{1,d}\!$. This asymmetry is a weighted
average of $A^{K^+}_{1,d}$ and $A^{K^-}_{1,d}$ with weights given by the spin-averaged
$K^+$ and $K^-$ cross-sections
\begin{equation}
A^{K^++K^-}_{1,d} = \frac{\sigma^{K^+} A^{K^+}_{1,d} + \sigma^{K^-} A^{K^-}_{1,d}}
{\sigma^{K^+} + \sigma^{K^-}}.
\label{asum}
\end{equation}
It is found to be very stable with respect to  the
ratio $\sigma^{K^-}/\sigma^{K^+}$.
Indeed a change of this ratio by $\pm 10 \%$  does not
modify  $A^{K^++K^-}_{1,d}\!$ by more than 10\% of its statistical error.
At LO, the cross-section ratio  only depends on the unpolarised PDFs and on the ratios
of unfavoured to favoured and strange to favoured FFs:  
\begin{eqnarray}
R_{UF} = \frac{ \int D_d^{K^+}(z) dz}{\int D_u^{K^+}(z) dz},
~~~
R_{SF} = \frac{ \int D_{\overline s}^{K^+}(z) dz}{\int D_u^{K^+}(z) dz}
\label{d321}
\end{eqnarray}
which are respectively equal to 0.13 and 6.6 for the DSS FFs at 
$Q^2 = 3\,({\rm GeV}/c)^2$ (0.35 and 3.4 for the EMC FFs).%
\footnote {These values remain practically unchanged 
when the range of $z$ is limited to 0.85 instead of 1.}
The values shown in Fig.\,\ref{fig:akpmchi} have been  obtained
with the MRST PDFs and the DSS FFs.
As for the $K^+$ and $K^-$ asymmetries, they are in  very good agreement 
with the HERMES values of Ref.\,\cite{hrm2005}.

For an isoscalar target, the charged kaon asymmetry and the inclusive asymmetry 
can be written at LO as
\begin{eqnarray}
A^{K^++K^-}_{1,d}\!\! = \xi  \frac {\Delta Q + \alpha \Delta s}{Q + \alpha s},
~~~
A_{1,d} = \xi \frac {\Delta Q +\frac{4}{5} \Delta s}{Q + \frac{4}{5} s}   
\label{aka}
\end{eqnarray}
where $Q$ is the  non-strange quarks density
$ Q = u + \overline u + d +\overline d$,
 the corresponding helicity density is $\Delta Q$, 
and  $\alpha = (2 R_{UF} + 2 R_{SF})/(2 + 3 R_{UF})$.

\begin{figure}[tb]
\begin{center}
\includegraphics[width=0.9\hsize,clip]{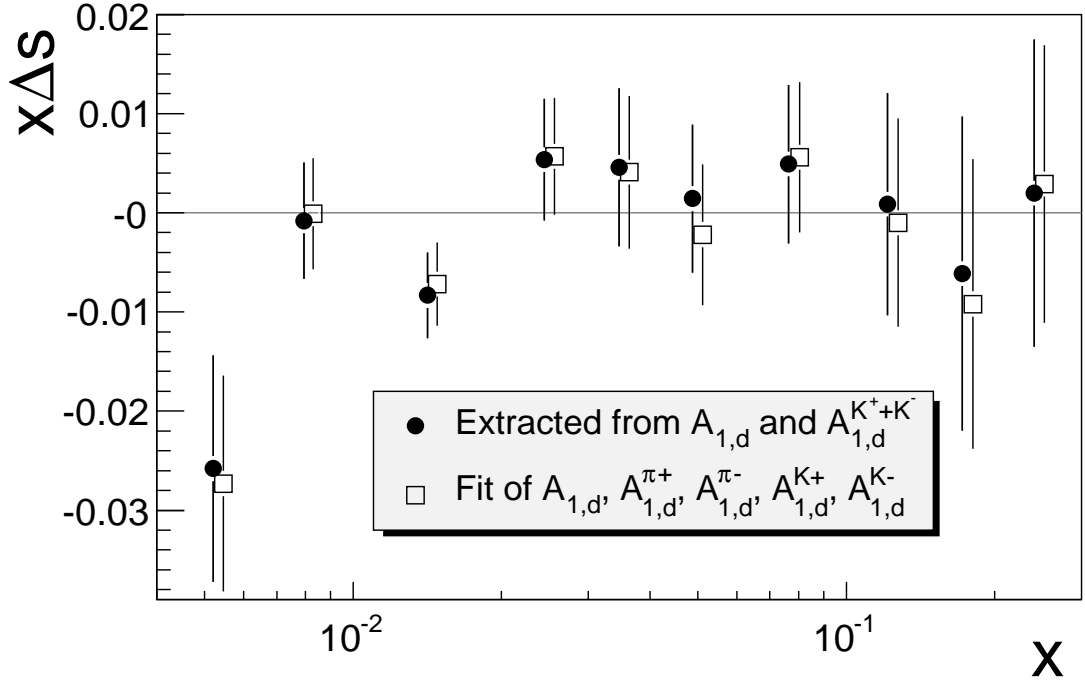}
\hfill
\end{center}
\caption{
   The strange quark spin distribution $x\, \Delta s(x)$ at $Q^2 = 3\,({\rm GeV}/c)^2$
  derived from the charged kaon asymmetry $A^{K^++K^-}_{1,d}$ 
  using DSS FFs and from $A_{1,d}$,
  compared to the result of the corresponding least square fit.
  The quoted errors are statistical only.
}
\label{fig:ds_xds}
\end{figure}

To take advantage of the  similarity between $A^{K^++K^-}_{1,d}$ and $A_{1,d}$,
it is convenient to write the strange quark polarisation $\Delta s/s$ in the form 
\begin{equation}
\frac {\Delta s}{s}  = \frac{1}{\xi} \Bigl [ A_{1,d} + (A^{K^++K^-}_{1,d} \!\! - A_{1,d}) \frac{Q/s + \alpha}{\alpha - 0.8} \Bigr ],
\label{ds_s}
\end{equation}
where $Q$ and $s$ are spin-independent non-strange and strange quark densities.
As expected, the use of this formula leads to values of $\Delta s$ 
practically equal to those of the least square fit but with slightly larger 
statistical errors (Fig.\,\ref{fig:ds_xds}).
The above formula shows that in the  special case where $A^{K^++K^-}_{1,d}\!\!$
would be strictly equal to $ A_{1,d}$,
the strange quark helicity would become insensitive to  FFs and its first moment 
would be  small 
and positive ($\approx 0.006$).
Otherwise the main dependence on the FFs is due to $R_{SF}$, which appears only
in the numerator of $\alpha$, and its effect is amplified by the large values 
of the ratio $Q/s$.
Negative values of $\Delta s$ correspond to negative values of  $A^{K^++K^-}_{1,d}$
at low $x$ where $A_1^d \approx 0$ and to $ A^{K^++K^-}_{1,d} \!< A_{1,d}$ at higher $x$.
The $A_{1,d}(x)$ prediction of the DSSV fit \cite{florian2008}
is shown in Fig.\,\ref{fig:akpmchi} for comparison. Neither the COMPASS nor
the HERMES points show any tendency to lie below the $A_{1,d}$ curve in the range
$0.03 < x < 0.3$. There is thus no indication for a significantly negative
$\Delta s$ in this region, in contrast to predictions of most
fits using only $g_1$ data, as shown 
 for instance by the DNS prediction in Fig.\,\ref{fig:PDF_Syst_EMC_DNS}.
 The COMPASS values of $A^{K^++K^-}_{1,d}$ provide at least
a hint that $\Delta s$ may become negative  in the previously unmeasured  low $x$ region
($x <0.02$), as predicted in the recent DSSV fit \cite{florian2008}.

\begin{figure}[tb]
\begin{center}
\includegraphics[width=0.9\hsize]{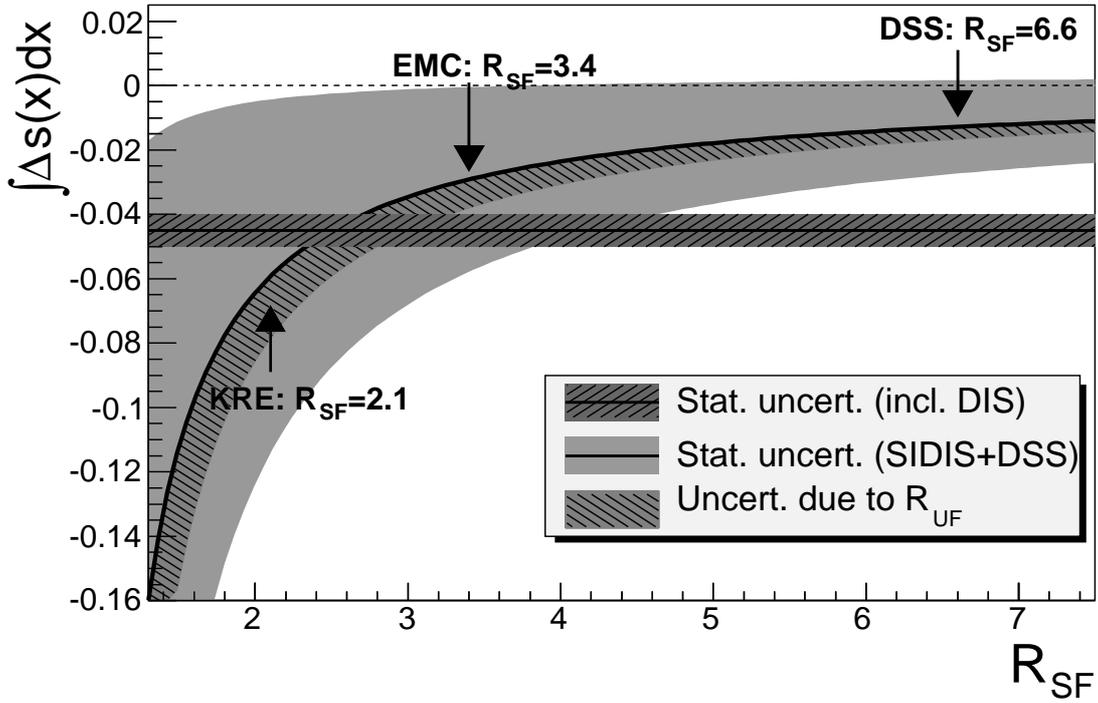}
\end{center}
\caption{
  Integral of $\Delta s$ over the measured range of $x$, as a function of the ratio
  $R_{SF}$ for $R_{UF}$ fixed at the DSS value of 0.13 (thick solid curve).
  The light-grey area shows the statistical uncertainty and the hatched band
  inside of it shows the effect of increasing $R_{UF}$ to 0.35 (EMC value).
  The horizontal band represents the full moment of $\Delta s$ derived from
  the COMPASS value of the first moment of $g_1^d(x)$ (Eq.\,\ref{mom_cmp}).
  The values of $R_{SF}$ corresponding to DSS\cite{florianFF},
 EMC\cite{EMC_FF} and KRE\cite{KRE} parameterisations 
  of FFs are indicated by arrows.
}
\label{fig:integ}
\end{figure}

Fig.\,\ref{fig:integ} shows the variation of the first moment of $\Delta s$
truncated to the measured region as a function of $R_{SF}$. For $R_{SF} \gtrsim 5$,
we observe that the values are close to zero and larger than the full moment 
derived from the inclusive analysis (Eq.\,\ref{mom_cmp}).
The contribution from the region $x > 0.3$ is limited by the positivity condition
$ |\Delta s(x)| \le s(x)$ and cannot exceed 0.003 in absolute value. Thus any difference
between the truncated SIDIS moment and the full moment must be compensated
by an unmeasured contribution in the low $x$ region. 
In particular this is the case for the DSS FFs where  $R_{SF} = 6.6$. 
The difference never exceeds two standard deviations, so that no firm conclusions can be 
drawn from the COMPASS data only, nevertheless, as shown on Fig.\,\ref{fig:akpmchi}, the
HERMES data lead to a similar result. In contrast, if  $R_{SF} \lesssim 4$, 
the asymmetry $A^{K^++K^-}_{1,d}$
becomes less and less sensitive to $\Delta s$ because $D_{\overline s}^{K^+}$ is small.
This is the case for the EMC FFs and, in general, for older parameterisations such as KRE \cite{KRE}.

%% file: conclusion.tex
\section {Conclusions}

We have presented a first measurement of the  longitudinal spin asymmetries
for charged pions and kaons identified with the RICH detector in the COMPASS
experiment. These measurements are used in combination with the inclusive
asymmetries to evaluate the polarised valence, non-strange sea and 
strange quark  distributions.
The results for valence quarks and non-strange sea quarks are in good agreement with
the DNS parameterisation. They show  weak dependence on the selected
parameterisation of the fragmentation functions.
 The distribution of $\Delta s$ is compatible with zero
in the whole measured range, in contrast to  the  shape of the strange quark
helicity distribution obtained in most LO and  NLO QCD fits.
The value
of the first moment of $\Delta s$ and its error are very sensitive to the assumed value of 
the ratio of the ${\overline s}$-quark to $u$-quark fragmentation functions into 
positive kaons 
$\int D_{\overline s}^{K^+}(z) dz/\int D_u^{K^+}(z) dz$.

%% file: acknowledgement.tex
\section*{Acknowledgements} 

We gratefully acknowledge the support of the CERN management and staff and 
the skill and effort of the technicians of our collaborating institutes. 
Special thanks go to V.~Anosov and V.~Pesaro for their technical support 
during the installation and the running of this experiment. 
This work was made possible thanks to the financial support of our funding agencies.

%% file: tables_withRC2.tex
\begin{table}[]
\caption{Values of the inclusive asymmetry $A_{1,d}$ with their
statistical and systematic errors as a function of $x$, 
with the corresponding average value of $Q^2$.}
\label{tab:DIS_asym}

\vspace*{3mm}
\begin{tabular}{|c|c|r|r|}
\hline
 $x$-bin & $\langle x\rangle$ &$\langle Q^2\rangle$~~~~ &$A_{1,d}\pm \delta A_{stat}\pm \delta A_{syst}$ \\
         &                    &   (GeV/$c)^2$   &                   \\
\hline
0.004--0.006 & 0.0052 &  1.17~~~~ & $ 0.001\pm0.005\pm0.002$ \\
0.006--0.010 & 0.0079 &  1.48~~~~ & $-0.001\pm0.003\pm0.001$ \\
0.010--0.020 & 0.0141 &  2.15~~~~ & $-0.002\pm0.003\pm0.001$ \\
0.020--0.030 & 0.0244 &  3.23~~~~ & $ 0.010\pm0.005\pm0.002$ \\
0.030--0.040 & 0.0346 &  4.33~~~~ & $ 0.003\pm0.006\pm0.003$ \\
0.040--0.060 & 0.0487 &  5.87~~~~ & $ 0.016\pm0.006\pm0.003$ \\
0.060--0.100 & 0.0765 &  8.63~~~~ & $ 0.039\pm0.007\pm0.004$ \\
0.100--0.150 & 0.121~ & 12.9~~~~~ & $ 0.090\pm0.010\pm0.008$ \\
0.150--0.200 & 0.172~ & 17.8~~~~~ & $ 0.126\pm0.015\pm0.011$ \\
0.200--0.300 & 0.240~ & 24.9~~~~~ & $ 0.159\pm0.017\pm0.014$ \\
\hline
\end{tabular}
\end{table}

\begin{sidewaystable}[]
\caption{Unfolded hadron asymmetries of charged pions and kaons.}
\label{tab:SIDIS_asym}

\vspace*{3mm}
\begin{tabular}{|c|r|r|r|r|r|}
\hline
$\langle x\rangle$ &$\langle Q^2\rangle$~~~~ &$A_{1,d}^{\pi+}\pm \delta A^{\pi+}_{stat}\pm \delta A^{\pi+}_{syst}$ & $A_{1,d}^{\pi-}\pm \delta A^{\pi-}_{stat}\pm \delta A^{\pi-}_{syst}$ & $A_{1,d}^{K+}\pm \delta A^{K+}_{stat}\pm \delta A^{K+}_{syst}$ & $A_{1,d}^{K-}\pm \delta A^{K-}_{stat}\pm \delta A^{K-}_{syst}$ \\
                   &   (GeV/$c)^2$    &      &   &   &  \\            
\hline
0.0052 &  1.16~~~~ & $  0.006\pm0.014\pm0.006$ & $  0.009\pm0.014\pm0.006$ & $ -0.018\pm0.029\pm0.011$ & $ -0.084\pm0.030\pm0.014$ \\
0.0079 &  1.42~~~~ & $ -0.003\pm0.008\pm0.003$ & $ -0.002\pm0.008\pm0.003$ & $ -0.017\pm0.017\pm0.007$ & $  0.014\pm0.018\pm0.007$ \\
0.0141 &  2.03~~~~ & $ -0.003\pm0.007\pm0.003$ & $ -0.007\pm0.007\pm0.003$ & $ -0.039\pm0.014\pm0.006$ & $ -0.004\pm0.016\pm0.006$ \\
0.0244 &  3.19~~~~ & $ -0.001\pm0.011\pm0.004$ & $  0.006\pm0.012\pm0.005$ & $  0.020\pm0.022\pm0.009$ & $  0.025\pm0.026\pm0.010$ \\
0.0346 &  4.43~~~~ & $  0.026\pm0.015\pm0.006$ & $  0.004\pm0.016\pm0.006$ & $  0.021\pm0.030\pm0.012$ & $  0.012\pm0.035\pm0.014$ \\
0.0487 &  6.10~~~~ & $  0.016\pm0.015\pm0.006$ & $  0.037\pm0.016\pm0.007$ & $  0.066\pm0.029\pm0.011$ & $ -0.058\pm0.035\pm0.015$ \\
0.0763 &  9.26~~~~ & $  0.046\pm0.017\pm0.008$ & $  0.018\pm0.018\pm0.007$ & $  0.064\pm0.032\pm0.013$ & $  0.015\pm0.041\pm0.017$ \\
0.121~ & 14.9~~~~~ & $  0.094\pm0.025\pm0.012$ & $  0.087\pm0.028\pm0.013$ & $  0.117\pm0.046\pm0.019$ & $ -0.007\pm0.065\pm0.026$ \\
0.171~ & 22.4~~~~~ & $  0.102\pm0.039\pm0.017$ & $  0.132\pm0.044\pm0.020$ & $  0.116\pm0.070\pm0.028$ & $  0.002\pm0.103\pm0.041$ \\
0.240~ & 32.8~~~~~ & $  0.218\pm0.044\pm0.024$ & $  0.147\pm0.051\pm0.023$ & $  0.208\pm0.078\pm0.031$ & $ -0.018\pm0.118\pm0.047$ \\
\hline
\end{tabular}
\end{sidewaystable}

\begin{sidewaystable}[]
\caption{Correlation coefficients of unfolded asymmetries in bins of $x$.}
\label{tab:corr}

\vspace*{3mm}
\begin{tabular}{|c||c|c|c|c|c|c|c|c|c|c|}
\hline
$x$-bin & $\!\!$0.004-0.006$\!\!$ &$\!\!$0.006-0.01$\!\!$ &$\!$0.01-0.02$\!$ &$\!$0.02-0.03$\!$ &$\!$0.03-0.04$\!$ &$\!$0.04-0.06$\!$ &$\!$0.06-0.10$\!$ &$\!$0.10-0.15$\!$ &$\!$0.15-0.20$\!$ &$\!$0.20-0.30$\!$\\
\hline
\hline
$A_{1,d}^{\pi+}\&A_{1,d}$        &  0.29 &  0.34 &  0.39 &  0.40 &  0.40 &  0.41 &  0.41 &  0.40 &  0.39 &  0.39 \\
\hline
$A_{1,d}^{\pi-}\&A_{1,d}$        &  0.30 &  0.35 &  0.38 &  0.39 &  0.39 &  0.39 &  0.38 &  0.36 &  0.35 &  0.34 \\
$A_{1,d}^{\pi-}\&A_{1,d}^{\pi+}$ &  0.12 &  0.15 &  0.17 &  0.16 &  0.16 &  0.16 &  0.16 &  0.15 &  0.15 &  0.17 \\
\hline
  $A_{1,d}^{K+}\&A_{1,d}$        &  0.11 &  0.15 &  0.17 &  0.19 &  0.19 &  0.20 &  0.21 &  0.21 &  0.21 &  0.21 \\
  $A_{1,d}^{K+}\&A_{1,d}^{\pi+}$ &$-0.18$&$-0.12$&$-0.08$&$-0.06$&$-0.05$&$-0.04$&$-0.04$&$-0.03$&$-0.02$&$-0.02$\\
  $A_{1,d}^{K+}\&A_{1,d}^{\pi-}$ &  0.03 &  0.04 &  0.04 &  0.04 &  0.04 &  0.04 &  0.04 &  0.05 &  0.05 &  0.05 \\
\hline
  $A_{1,d}^{K-}\&A_{1,d}$        &  0.11 &  0.14 &  0.16 &  0.16 &  0.16 &  0.16 &  0.16 &  0.15 &  0.14 &  0.14 \\
  $A_{1,d}^{K-}\&A_{1,d}^{\pi+}$ &  0.03 &  0.03 &  0.03 &  0.03 &  0.03 &  0.03 &  0.03 &  0.03 &  0.03 &  0.03 \\
  $A_{1,d}^{K-}\&A_{1,d}^{\pi-}$ &$-0.14$&$-0.09$&$-0.05$&$-0.04$&$-0.04$&$-0.03$&$-0.03$&$-0.03$&$-0.03$&$-0.04$\\
  $A_{1,d}^{K-}\&A_{1,d}^{K+}$   &  0.05 &  0.07 &  0.09 &  0.10 &  0.10 &  0.10 &  0.10 &  0.10 &  0.11 &  0.12 \\
\hline
\end{tabular}
\end{sidewaystable}
